\documentclass[twocolumn,pra,aps,superscriptaddress,showpacs]{revtex4}
\usepackage{graphicx}
\usepackage{bm}

\begin{document}

\title{Understanding destruction of $n$th-order quantum coherence in terms of
multi-path interference}
\author{D.L. Zhou}
\affiliation{Center for Advanced Study, Tsinghua University, Beijing 100084, China}
\affiliation{Institute of Theoretical Physics, Academia Sinica, \\
P.O.Box 2735, Beijing 100080, China}
\author{P. Zhang}
\affiliation{Institute of Theoretical Physics, Academia Sinica, \\
P.O.Box 2735, Beijing 100080, China}
\author{C.P. Sun}
\email{suncp@itp.ac.cn}
\homepage{http:// www.itp.ac.cn/~suncp}
\affiliation{Institute of Theoretical Physics, Academia Sinica, \\
P.O.Box 2735, Beijing 100080, China}

\begin{abstract}
The classic example of the destruction of interference fringes in a
``which-way'' experiment, caused by an environmental interaction, may be
viewed as the destruction of first-order coherence as defined by Glauber
many years ago \cite{Glauber}. However, the influence of an environment can
also destroy the $n$th-order quantum coherence in a quantum system, where
this high order coherence is captured. We refer to this phenomenon as the $n$%
th-order decoherence. In this paper we show that, just as the
first-order coherence can be understood as the interference of the
amplitudes for two distinct paths, the higher order coherence may
be understood as the interference of multiple amplitudes
corresponding to multiple paths. To see this, we introduce the
concept of $n$th-order ``multi-particle wave amplitude''. It turns
out that  the $n$th-order correlation function can be expressed as
the square norm of some ``multi-particle wave amplitude'' for the
closed system or as the sum of such square norms for the open
system. We also examine, as a specific example, how an environment
can destroy the second order coherence by eliminating the
interference between various multiple paths.
\end{abstract}

\maketitle




\section{Introduction}

A most profound concept in quantum physics is quantum coherence. The first
order version of quantum coherence can be directly manifested by the
superposition of two quantum states. This is quite similar to the optical
coherence in Young's double experiment . From the standpoint of the photon-
detection theory by Glauber, this simple coherence can be mathematically
depicted by the first order correlation function \cite{Glauber}. However, in
quantum mechanics, this first order coherence phenomenon does not sound very
marvellous since the same circumstance can also occur in a classical case,
such as the optical interference in the above mentioned Young's double
experiment. In fact, by the the first order correlation function only, it is
impossible to distinguish the natures of a laser light field and a
conventional light field with identical spectral properties. As an effective
remedy, Glauber's $n$-th order quantum correlation function, which accounts
for various intrinsic ( $n$-th order) quantum coherence effects, such as the
intensity-intensity correlation measurement in the Hanbury-Brown-Twiss
experiment\cite{Twiss}, was introduced \cite{Glauber}. Indeed, this function
reflects the intrinsically quantum features of coherence beyond the
classical analogue.

The quantum coherence embodies the wave nature in the world of microscopic
particles. On the other hand, it is very fragile and can easily be destroyed
by a ``which-path(way)'' experiment \cite{Rampe,Umansky}, or by an
environmental interaction. This phenomenon of destruction of coherence is
usually referred to as quantum decoherence \cite{Zurek,Haroche}. The
phenomenon of losing the coherence described by the first order correlation
function is defined as the first order quantum decoherence. Motivated by the
considerations in fundamental quantum measurement problem \cite{Sun-qm,Knight}%
, and also by the attempts to preserve quantum coherence of qubits in
quantum computing \cite{Unruh,Palma,Sun-qc}, many recent experimental and
theoretical investigations have been focused on revealing the physical
mechanism of the decoherence problem, e.g., see \cite{Sun-adia}. According to
these studies, this first order decoherence can be roughly understood
through the quantum entanglement of the considered system with the
environment or the measuring apparatus.

Obviously, this entanglement implies a ``which-path(way)'' detection \cite
{Rampe} in the single particle picture. Precisely speaking, in an initial
coherent superposition $|\psi _s\rangle =\sum c_n|n\rangle $ , each system
state $|n\rangle $ corresponds to a ``path'' and many ``two path''
interferences are reflected in the square norm of the spatial wave function $%
\langle x|\psi_s\rangle $.Thus by considering $\langle x|\psi_s\rangle $
the quantum coherence can be captured to some extent. After the interaction,
each ``path'' is correlated with an environment state $|e_n\rangle $ to form
an entangling state $|\psi_T\rangle =\sum c_n|n\rangle \bigotimes
|e_n\rangle $. Here, the different states $|e_n\rangle $ distinguish among
the ``paths''of different $|n\rangle $ and thus record the information of
each ``path''. The interference terms in the spatial intensity $%
I(x)=Tr(\langle x|\psi _T\rangle \langle \psi _T|x\rangle )$ will disappear
when the environment states $|e_n\rangle $ are completely distinct, i.e., $%
\langle e_m|e_n\rangle =\delta _{mn}$. In that case each path is labeled by
an environment state .

The above well-known explanation of the first order decoherence in terms of
``which-path(way)'' detection mechanism is simple but very profound.
However, it is not yet clear whether this mechanism can be used to elucidate
the $n$th order quantum decoherence (n-QDC), the destruction of quantum
coherence described by Glauber's $n$th order correlation function. The
difficulty is we do not exactly know what are the ``paths'' and the
corresponding ``which-path'' detection. Most recently, we have touched the
second order quantum decoherence problem \cite{Zhou-dl} bypassing this
difficulty. As a matter of fact, in our treatment, we did not define the
concept of ``path'' directly. The concrete calculation in the ref. \cite
{Zhou-dl} motivated us to consider the ``which-path'' picture of the higher
order quantum decoherence in general.

In our present investigation, it is crucial to show, we notice
that, for a close system, in some cases Glauber's $n$th-order
correlation function is the square norm of the $n$th-order
``multi-particle wave amplitude'', which will be defined later in
this paper, while for an open system, it can become a sum of the
square norms of the $n$th-order ``multi-particle wave amplitude''
over the states of an environment or an apparatus interacting with
this open system. This observation is crucial in our present
investigation. As an effective wave function, this multi-particle
amplitude can be shown to be a supposition of many generalized
``paths'' (the multi- particle paths or simply multiple paths).
Thus, the higher order coherence may be understood as the
interference of multiple particle amplitudes. With this conception
a generalized ``which-path'' detection may be established in terms
of ``multi-particle paths'' as the physical mechanism of
higher-order decoherence for some examples.

In section $2$, we will briefly explain Glauber's $n$th order quantum
coherence in terms of the single and multi photon effective wave functions
used in \cite{Scully,Shih}. As their generalizations, in section 3 the
concepts of multi-particle path and multi-particle wave amplitude are
introduced for both close and open bosonic systems. Especially, the $n$th
order correlation functions of bosonic systems will be studied. In section
4, an interacavity model with two bosonic modes is used to demonstrate $2$%
-QDC as a ``which-path'' detecting process. In section 5, the exact solution
obtained in the appendix is utilized to show the dynamical process of $2$%
-QDC , which is caused by the entanglement with the environment or an
apparatus and indeed can be explained as a generalized ``which-path''
measurement for the explicitly -defined multi-particle paths.

\section{The $n$th-Order Coherence for Quantized Light Field and
Multi-Photon Wave Amplitude}

In quantum mechanics, a pure state is a superposition $|\psi \rangle
=\sum_kc_k|k\rangle $ of many components $|k\rangle $ , but the
corresponding mixture $\rho =\sum_k|c_k|^2|k\rangle \langle k|$ can describe
the same classical probability distribution $|c_k|^2.$ However, $|\psi
\rangle $ and $\rho $ represents different quantum realities. Usually it is
said that the two components of a pure state is more coherent than those of
a mixed one. This coherence property is obviously reflected by the intensity
interference of two ``paths'' corresponding to the two components in the
quantum state. For the system of one particle , only single particle
property is relevant for this observation. In this sense, only the intensity
interference experiment is essential for one particle system. For many
particle system, however, there exist many experiments (such as the
Hanbury-Brown-Twiss experiment and the intensity-intensity correlation
measurement) to show the much richer natures of quantum coherence.

In order to study quantum coherence in many particle system,
Glauber introduced the so-called $n$-th order quantum correlation
function ($n$-QCF)
\begin{widetext}
\begin{equation}
G^{(n)}[\alpha _1,t_1;\alpha _2,t_2;\cdots ;\alpha _n,t_n]
=Tr[\hat{\rho}E_{\alpha _1}^{\dagger }(t_1)E_{\alpha _2}^{\dagger
}(t_2)\cdots E_{\alpha _n}^{\dagger }(t_n)E_{\alpha _n}(t_n)\cdots
E_{\alpha _2}(t_2)E_{\alpha _1}(t_1)],
\end{equation}
\end{widetext}
for the electro-magnetic field $E_\alpha (t)$ in different mode $\alpha$.
Here, $E_\alpha (t)$ is the annihilation operator of mode $\alpha $ at time $%
t$ in the Heisenberg picture, $E_\alpha ^{\dagger }(t)$ is the
corresponding conjugate operator, and the density matrix
$\hat{\rho}$ represents the initial state of many-mode
electro-magnetic field. Of course, this formalism can also be used
to study the coherence property for any quantum many-body system.

Furthermore, to describe the higher order coherence, Glauber also defined
the $n$th-order coherence function in the form \cite{footnote}
\begin{widetext}
\begin{equation}
g_{\alpha \beta } \equiv g[\alpha _1,\alpha _2,\cdots ,\alpha
_n;\beta _1,\beta _2,\cdots ,\beta _n]
=\frac{Tr[\hat{\rho}E_{\beta _1}^{\dagger }(t_1)E_{\beta
_2}^{\dagger }(t_2)\cdots E_{\beta _n}^{\dagger }(t_n)E_{\alpha
_n}(t_n)\cdots E_{\alpha _2}(t_2)E_{\alpha
_1}(t_1)]}{\sqrt{G[\alpha _1,t_1;\alpha _2,t_2;\cdots ;\alpha
_n,t_n]}\sqrt{G[\beta _1,t_1;\beta _2,t_2;\cdots ;\beta _n,t_n]}},
\end{equation}
\end{widetext}
Obviously, this coherence function is defined by a ratio of the
off-diagonal elements of the reduced multi-time density matrix
$g=(g_{\alpha \beta }:\alpha =(\alpha _1,\alpha _2,\cdots ,\alpha
),\beta =(\beta _1,\beta _2,\cdots ,\beta _n))$to its diagonal
ones. It is easily observed from this definition that these
off-diagonal elements represent the coherence effect, and each one
$g_{\alpha \beta }$ correlates two diagonal ones $g_{\alpha \alpha
}$and $g_{\beta \beta }.$ Thus, we can understand the coherence
function as measuring the degree of coherence for the two diagonal
elements, which correspond to two different ``paths''. From this
point of view, there is obviously no quantum coherence for a
completely mixed state with vanishing off-diagonal elements. In
the following discussion, we will give simple examples to
illustrate this viewpoint.

Now let us turn to an instance \cite{Scully} from which we can see clearly
how the higher order quantum coherence effects given by $2$-QCF is revealed
in the multi-particle picture. For a single photon, the coherent
superposition $|s\rangle =\frac 1{\sqrt{2}}(|1_k\rangle +|1_{-k}\rangle )$
of two states with opposite wave vectors $k$ and $k^{\prime }$ possesses the
first order quantum coherence which can be described by the interference
fringes
\begin{eqnarray}
G^{(1)}(r,r,t)&=&\langle
s|E^{-}(r,t)E^{+}(r,t)|s\rangle\nonumber\\
&=&|\langle 0|E^{+}(r,t)|s\rangle |^2\propto \cos ^2(kr),
\end{eqnarray}
where
\[
E^{+}(r,t)=\sum E_ka_k\exp (ikr-i\omega _kt)
\]
is the photon field operator with positive frequency for the annihilation
operator $a_k$ . It should be noticed that the diagonal element $%
G^{(1)}(r,r,t)$ of the first order correlation function is just the square
norm of the single photon wave-packet \cite{Shih}
\begin{eqnarray}
\langle 0|E^{+}(r,t)|s\rangle&=&\frac1{\sqrt{2}}\{\langle
0|E^{+}(r,t)|1_k\rangle+\langle 0|E^{+}(r,t)|1_k\rangle\}\nonumber \\
&=&\frac1{\sqrt{2}}(E_ke^{ikr-i\omega _kt}+E_{-k}e^{-ikr-i\omega
_{-k}t})
\end{eqnarray}
Therefore, $G^{(1)}(r,r,t)$ represents the interference between
the ``two paths'' $\langle 0|E^{+}(r,t)|1_k\rangle $ and $\langle
0|E^{+}(r,t)|1_k\rangle .$ This just developes the corresponding
concept in Young's double experiment for quantized light field.
The above reformulation of the first order quantum coherence
implies that ``two paths'' are necessary for the interference
phenomenon.

If we consider the two particle state $|1_k,1_{k^{\prime }}\rangle $ with a
single component, an interesting situation arises where the first order
quantum coherence does not appear, but we can see the second order effect
through the second order quantum correlation function
\begin{equation}
G^{(2)}(r_1,r_2,t_1,t_2)=2E_k^4\{1+\cos [(k-k^{\prime })(r_1-r_2)]\}.
\end{equation}
Unlike the case of first order coherence, in this case there do not appear
the obvious two or many ``paths''. Nevertheless, the interference phenomenon
can still be captured in a similar way with the introduction of generalized
``path''. We understand the generalized ``path'' as described by The ``two
time'' correlation function
$G^{(2)}(r_1,r_2,t_1,t_2)$ $=|\psi |^2$ where $\psi $ is the ``two-photon
wave function''
\begin{eqnarray}
\psi &\equiv &\psi (r_1,r_2,t_1,t_2)=\langle
00|E^{+}(r_2,t_2)E^{+}(r_1,t_1)|1_k,1_{k^{\prime }}\rangle  \nonumber \\
&=&E_k^2e^{-2i\omega _kt}[e^{ikr_1+ik^{\prime }r_2}+e^{ik^{\prime
}r_1+ikr_2}]
\end{eqnarray}
$\psi$ was invoked as a two photon effective wave-function, and it was also
called the biphoton wave packet for the photon field $E^{+}(r,t)$ \cite{Shih}%
. Especially, we remark that the biphoton wave packet $\psi $ is a coherent
superposition of two ``probability amplitudes'' corresponding to two
``two-photon paths''
\[
\langle 00|E^{+}(r_2,t_2)|0_k,1_{k^{\prime }}\rangle \langle
0_k,1_{k^{\prime }}|E^{+}(r_1,t_1)|1_k,1_{k^{\prime }}\rangle
\]
and
\[
\langle 00|E^{+}(r_2,t_2)|1_k,0_{k^{\prime }}\rangle \langle
1_k,0_{k^{\prime }}|E^{+}(r_1,t_1)|1_k,1_{k^{\prime }}\rangle
\]

Starting from Glauber's standpoint and proceeding along, we come to the
conclusion that a set $\{G^{(n)}[\alpha _1,t_1;\alpha _2,t_2;\cdots ;\alpha
_n,t_n]|n=1,2,...\}$of correlation functions, rather than a single one, is
indispensable to describe comprehensively the wave-particle dual nature in
the quantum world of many particle system. The above example of second order
coherence shows that the conception of two-path interference in single
photon picture still works with a proper generalization of the concept of
``path''. Hence it is quite natural to seek a generalized
``which-path(way)'' measurement as the mechanism of higher-order quantum
decoherence.

\section{Multi-Particle Amplitude for free Bosons}

In this section the conception of two photon effective wave-function will be
generalized. It will be applied to the study of general quantum systems of
identical particles. We first discuss the spatially-homogeneous case for the
sake of simplicity.

We consider a homogeneous bosonic field with two modes $|V\rangle $ and $%
|H\rangle $. The generalized field operator in time-domain
\begin{eqnarray}
\hat{\phi}(t)&=&c_V\hat{b}_Ve^{-i\omega _Vt}+c_H\hat{b}_He^{-i\omega
_Ht}\nonumber\\ &\equiv& c_V(t)\hat{b}_V+c_H(t)\hat{b}_H  \label{eq1}
\end{eqnarray}
is an annihilation operator with respect to the superposition state
\begin{equation}
|+\rangle =c_V^{*}|V\rangle +c_H^{*}|H\rangle.
\end{equation}
Here , $b_H$ and $b_V$ are the annihilation operators of the boson system ; $%
c_V$ and $c_H$ satisfy the normalization relation $|c_V|^2+|c_H|^2=1$.
Without loss of generality, we take $c_V=c_H=1/\sqrt{2}$ . This means we
consider the measurement to detect the polarized boson along the 45$^o$
direction in the $V-H$ plane. We call $\hat{\phi}$ a ``measuring'' operator .

Corresponding to $\hat{\phi}$, the generalized second order correlation
function \cite{Zhou-dl}
\begin{eqnarray}
{G}^{(2)} &=&\langle 1_V1_H|\hat{\phi}^{\dagger }(t_1)\hat{\phi}^{\dagger
}(t_2)\hat{\phi}(t_2)\hat{\phi}(t_1)|1_V,1_H\rangle  \nonumber \\
\mbox{} &=&|\langle 0,0|\hat{\phi}(t_2)\hat{\phi}(t_1)|1_V,1_H\rangle
|^2\equiv |\Psi (t_1,t_2)|^2.
\end{eqnarray}
The two time wave function
\[
\Psi (t_1,t_2)=\langle 0,0|\hat{\phi}(t_2)\hat{\phi}(t_1)|1_V,1_H\rangle
\]
can be understood in terms of the two ``paths'' picture from the initial
state $|1_V,1_H\rangle $ to the final state $|0,0\rangle $: \vskip 5mm

\begin{center}
\begin{tabular}{|c|}
\hline
$
\begin{array}{ccccc}
|1_V,1_H\rangle & \stackrel{c_H(t_1)}{\longrightarrow } & |1_V,0_H\rangle &
\stackrel{c_{_V}(t_2)}{\longrightarrow } & |0,0\rangle \\
& \searrow \stackrel{c_V(t_1)}{} &  & \stackrel{c_H(t_2)}{}\nearrow &  \\
&  & |0_V,1_H\rangle &  &
\end{array}
$ \\ \hline
\end{tabular}
\end{center}
The two ``paths'' are just associated with the two amplitudes forming a
coherent superposition
\begin{eqnarray}
\Psi (t_1,t_2)&=&c_Vc_H\exp (-i\omega _Vt_2-i\omega
_H)+\nonumber\\
&&c_Hc_V\exp (-i\omega _{_H}t_2-i\omega _{_V}t_1)
\end{eqnarray}
Correspondingly, the second order correlation function
\begin{equation}
G^{(2)}=2|c_Vc_H|^2\{1+\cos [(\omega _V-\omega _H)(t_2-t_1)\}
\end{equation}

The above discussion for the second order quantum coherence is applicable to
the higher order case. Our arguments in this paper are based on two novel
observations: $a.$ The generalized field operator $\hat{\phi}=\sum c_n\hat{b}%
_n$ is specified for a quantum measurement about a superposition single
particle state $|\phi \rangle =\sum c_n^{*}|n\rangle $. $b.$ For a certain
initial single component state $|s_0\rangle $ of $N$ particles system, the $%
n $-th order quantum correlation function
\begin{equation}
G^{(n)}(r_1,r_2,\cdots ,r_n,t_1,t_2,\cdots ,t_n)=|\psi ^{(n)}|^2
\end{equation}
can be written as the norm square of an effective wave function $\psi ^{(n)}$%
, which is just a superposition of many amplitudes.

Let us consider the third order situation as an example. Let the initial
state be $|1_H,2_V\rangle $. Then the generalized third order correlation
function
\begin{eqnarray}
{G}^{(3)} &=&\langle 2_V1_H|\hat{\phi}^{\dagger }(t_1)\hat{\phi}^{\dagger
}(t_2)\hat{\phi}^{\dagger }(t_3)\hat{\phi}(t_3)\hat{\phi}(t_2)\hat{\phi}%
(t_1)|2_V,1_H\rangle  \nonumber \\
\mbox{} &=&|\langle 0,0|\hat{\phi}(t_3)\hat{\phi}(t_2)\hat{\phi}%
(t_1)|2_V,1_H\rangle |^2\nonumber\\
&\equiv& |\Psi (t_1,t_2,t_3)|^2
\end{eqnarray}
is a norm square of the two time wave function $:$
\begin{eqnarray}
\Psi (t_1,t_2,t_3) &=&\langle 0,0|\hat{\phi}(t_3)\hat{\phi}(t_2)\hat{\phi}%
(t_1)|2_V,1_H\rangle  \nonumber \\
&=&\sqrt{2}{c_V}^2c_He^{-i\omega _V(t_3+t_2)-i\omega _Ht_1}+  \nonumber \\
&&\sqrt{2}c_H{c_V}^2e^{-i\omega _Ht_2-i\omega _V(t_3+t_1)}+\nonumber\\
&&\sqrt{2}c_H{c_V}%
^2e^{-i\omega _Ht_3-i\omega _V(t_2+t_1)}
\end{eqnarray}
Each term in the above effective wave function is contributed by the
corresponding one of the four ``paths'' from $|2_V,1_H\rangle $ to $%
|0,0\rangle $:
\begin{widetext}
\centering
\begin{tabular}{|l|}
\hline
\begin{tabular}{ccccccc}
&  & $|2_V,0_H\rangle $ & $\stackrel{c_{_V}(t_2)}{\longrightarrow }$ & $%
|1_V,0_H\rangle $ &  &  \\
& $\nearrow \stackrel{c_{_H}(t_1)}{}$ &  &  &  & $\stackrel{c_V(t_3)}{}%
\searrow $ &  \\
$|2_V,1_H\rangle $ & $\stackrel{c_{_V}(t_1)}{\longrightarrow }$ & $%
|1_V,1_H\rangle $ & $\stackrel{c_H(t_2)}{\longrightarrow }$ & $%
|1_V,0_H\rangle $ & $\stackrel{c_V(t_3)}{\longrightarrow }$ & $|0,0\rangle $
\\
&  &  & $\searrow \stackrel{c_V(t_2)}{}$ &  & $\stackrel{c_{_H}(t_3)}{}%
\nearrow $ &  \\
&  &  &  & $|0_V,1_H\rangle $ &  &
\end{tabular}
\\ \hline
\end{tabular}
\end{widetext}

In terms of the effective $3$-time-wave function defined above, the third
order correlation function is explicitly written down:
\begin{eqnarray}
G^{(3)} &=&4|{c_V}^2c_H|^2(\frac 32+\cos [(\omega _V-\omega _H)(t_2-t_1)]+ \nonumber\\
&&\cos [(\omega _V-\omega _H)(t_3-t_1)]+\cos [(\omega _V-\omega
_H)(t_2-t_3)])  \nonumber
\end{eqnarray}
It shows the quantum interference in the time-domain.

The above analysis is valid only for the case where the considered system is
isolated from an environment and not measured by a detecting apparatus-a
detector. For our purpose, we need to consider an open system $S$
interacting with an environment (reservoir) or a detector $E$, and we must
extend the concepts of multi-particle(time)-wave functions and the
corresponding many-particle paths defined above. To do the generalization ,
we first invoke the effective field operator
\begin{eqnarray}
\hat{B}(t) &=&U^{\dagger }(t,0)\hat{\phi}(0)U(t,0)  \nonumber \\
&=&\exp (i\hat{V}t)\hat{\phi}(t)\exp (-i\hat{V}t)
\end{eqnarray}
where $\hat{\phi}(0)=c_V\hat{b}_V+c_H\hat{b}_H$ has been given by
Eq.(\ref{eq1}).
Then instead of the free time evolution governed by the free Hamiltonian $%
H_0 $, we use the evolution operator $U(t,0)$ governed by the total
Hamiltonian
\[
H=H_0+W+H_E\equiv H_0+V
\]
taking into account the role of the interaction between $E$ and $S$ . Here, $%
H_E$ is the free Hamiltonian for $E.$ If we only consider an ideal quantum
decoherence process without dissipation, $V$ should possess the nature of
quantum non- demolition: $[H_0,W]=0$ \cite{Sun-qm,Sun-adia}.

Let the states $|n\rangle $ $\equiv $ $|n_V,n_H\rangle $ be the common
eigen-states of $H_0$ and $W$ corresponding to the egen-vaslues $E_n$ and $%
V(n)(n=(n_V,n_H)).$ If the initial state of the total system is
\begin{equation}
|\psi (0)\rangle =|\phi _s\rangle \otimes |\phi _E\rangle
\end{equation}
where $|\phi _s\rangle $ and $|\phi _E\rangle $ are some specially-given
initial states of $S$ and $E$ respectively, we can define the effective
two-time state vector
\begin{equation}
|\psi _B(t,t^{\prime })\rangle =\hat{B}(t^{\prime })\hat{B}(t)|\psi
(0)\rangle .
\end{equation}
as a reasonable generalization of the effective ``two-time wave function''
given above. Its norm is just the second order correlation function
\begin{eqnarray}
\langle \psi _B(t,t^{\prime })|\psi _B(t,t^{\prime })\rangle &=&Tr(\hat{\rho}%
(0)\hat{B}^{\dagger }(t)\hat{B}^{\dagger }(t^{\prime })\hat{B}(t^{\prime })%
\hat{B}(t))  \nonumber \\
&=&G^{(2)}[t,t^{\prime },\hat{\rho}(0)],
\end{eqnarray}
for the density matrix $\hat{\rho}(0)=|\psi (0)\rangle \langle \psi (0)|.$

In fact, due to the non- demolition interaction not resulting in
dissipation, the basic dynamic properties of the open system do
not change even in the presence of $E$ . If we choose $|\phi
_s\rangle =|1_H,1_V\rangle ,$ there are only two ``paths'' from
the initial state $|1_V,1_H\rangle $ to the final state
$|0,0\rangle $ for $S$ , and $|0,0\rangle $ is the unique
state which can be reached by the action of $\hat{B}(t^{\prime })\hat{B}(t).$
Then,
\begin{widetext}
\begin{eqnarray}
\langle \psi _B(t,t^{\prime })|\psi _B(t,t^{\prime })\rangle
&=&\sum_{n,\beta }\langle \psi _B(t,t^{\prime })|n,\beta \rangle \langle
n,\beta |\psi _B(t,t^{\prime })\rangle =\sum_\beta |\langle 0,\beta |\psi
_B(t,t^{\prime })\rangle |^2  \nonumber \\
&=&\sum_\beta |\langle 0,\beta |\hat{B}(t^{\prime })\hat{B}(t))|1_H,1_V,\phi
_E\rangle |^2=\sum_\beta |\Psi _\beta (t_1,t_2)|^2
\end{eqnarray}
\end{widetext}
Here, the summation ranges over the complete set of states $|\beta \rangle $
of $E,$ and each term in the sum is a norm square of the effective two
particle wave function
\begin{equation}
\Psi _\beta (t_1,t_2)=\langle 0,\beta |\hat{B}(t^{\prime })\hat{B}%
(t))|1_H,1_V,\phi _E\rangle
\end{equation}
for the open system.

From the above calculations for the second and third order quantum
decoherence, we observe that for a specially-given initial state,
a higher order correlation function may be explicitly written down
as the norm square (or its sum ) of the multi-time-wave function,
which is a coherent superposition of several complex components
associated with the generalized many-particle paths. It is pointed
out that this kind of many-particle path is not a simple-product
of single-particle paths, but it can be determined by the
specially designed measurement.

\section{Generalized Which-Path Detection in An Intracavity Model}

In this section, an intracavity model is presented to demonstrate
in multi-particle picture the ``which-path'' detection
associated with higher-order quantum decoherence .

In a recent paper \cite{Zhou-dl}, we have studied the problem of $2$-QDC for
a cavity-QED system. The concrete calculation in the ref.\cite{Zhou-dl}
shows that the $2$-QDC effects can indeed be observed in the proposed
experiment. But it involves dissipation effect losing energy. However, it is
well known that quantum decoherence can still occur for an energy conserving
system. So in principle dissipation is not indispensible for the discussion
about decoherence effect. For this reason it is natural and interesting to
consider pure decoherence process without dissipation. The pure decoherence
can be well understood through the quantum entanglement of the considered
system with the environment or the measuring apparatus. For a model with
pure decoherence process, many concepts (such as multi-particle path and the
corresponding which-path detection ) can be made much clearer. Unlike the
approximately-solvable model treated in the ref.\cite{Zhou-dl}, which loses
its energy and coherence simultaneously, the model proposed in this section,
a bosonic system of two modes interacting with an external system of many
harmonic oscillators, does not dissipate its energy. This property makes the
model exactly solvable, and as a result the problem of higher-order quantum
decoherence can be studied in a straightforward way.

By taking $\hbar =1,$ the model Hamiltonian
$\hat{H}=\hat{H}_0+\hat{V}$ is defined by
\begin{eqnarray}
\hat{H}_0 &=&\omega _V\hat{b}_V^{\dagger }\hat{b}_V,   \\
\hat{V} &=&\sum_j\omega _j\hat{a}_j^{\dagger }\hat{a}_j+\sum_j[d_V(\omega _j)%
\hat{b}_V^{\dagger }\hat{b}_V+d_H(\omega _j)\hat{b}_H^{\dagger }\hat{b}_H]\nonumber\\
&&\hat{a}_j^{\dagger }+\hat{a}_j),
\end{eqnarray}
where $\hat{H}_0$ is the free Hamiltonian of the system, $\hat{V}$ the free
Hamiltonian $\sum_j\omega _j\hat{a}_j^{\dagger }\hat{a}_j$ of the reservoir
(or a detector) plus a non-demolition interaction between the system and the
reservoir; and $\hat{b}_V^{\dagger }(\hat{b}_V),\hat{b}_H^{\dagger }(\hat{b}%
_H)$ the creation (annihilation) operators for the two modes with
frequencies $\omega _V$ ($\neq 0$) and $\omega _H=0$. The operators $\hat{a}%
_j^{\dagger }(\hat{a}_j)$ are creation (annihilation) operators of the
reservoir modes of the frequencies $\omega _j$. The frequency-dependent
constant $d_H(\omega _j)$ ($d_V(\omega _j)$) measures the coupling constant
between $H$ ($V$) mode and the $j$ mode of the reservoir.

The most important feature of the model is the non-demolition condition $%
[H_0,V]=0$. It means that the system does not dissipate energy to the
reservoir. On the other hand, the system can leave imprint on the reservoir
since, for different number states $|n_V,n_H\rangle $, there are different
interactions
\[
\sum_j[n_Vd_V(\omega _j)+n_Hd_H(\omega _j)](\hat{a}_j^{\dagger }+\hat{a}_j)
\]
acting on the oscillator reservoir with different driving forces $\sim
n_Vd_V(\omega _j)+n_Hd_H(\omega _j)$. When there is only one mode in the
external system(reservoir) the whole system can physically be described by
an intracavity model: Two mode cavity field interact with a moving wall of
the cavity, which is attached to a spring and can be regarded as a harmonic
oscillator with a small mass [4,8]. The fields are coupled to the cavity
wall (a moving mirror) by the radiation pressure forces in proportion to the
photon numbers $\hat{b}_H^{\dagger }\hat{b}_H$ and $\hat{b}_V^{\dagger }\hat{%
b}_V$.

For the above introduced model, we can discuss the higher order
decoherence problem in the Heisenberg picture by explicitly
defining the many-particle ``which -path'' measurement. The second
order coherence is directly determined by the second order
correlation function.
\begin{equation}
G[t,t^{\prime },\hat{\rho}(0)]=Tr(\hat{\rho}(0)\hat{B}^{\dagger }(t)\hat{B}%
^{\dagger }(t^{\prime })\hat{B}(t^{\prime })\hat{B}(t))
\end{equation}
which is defined as a functional of the density operator $\hat{\rho}(0)$ of
the whole system at a given time $0$. Here, the bosonic field (measuring)
operator
\begin{equation}
\hat{B}(t)=\exp (i\hat{V}t)(c_H\hat{b}_H+c_V\hat{b}_V\exp (-i\omega
_Vt))\exp (-i\hat{V}t)
\end{equation}
is defined for the interacting system. Like the operator defined
for the non-interacting system, it also describes a specific
destructive quantum measurement [4] with respect to the polarized
states
\[
|+\rangle =c_H^{*}|H\rangle +c_V^{*}|V\rangle
\]
where $c_H$ and $c_V$ satisfy the normalization relation $|c_H|^2+|c_V|^2=1$%
. Without loss of the generality, we take $c_H=c_V=1/\sqrt{2},$ considering
a specific measuremeant.

To examine whether the macroscopic feature of the reservoir causes the
second order decoherence or not , we consider the whole system in an initial
state
\begin{equation}
|\psi (0)\rangle =|1_H,1_V\rangle \otimes |\{0_j\}\rangle ,
\end{equation}
where $|\{0_j\}\rangle$ is the vacuum state of the reservoir. Here, we have
denoted the general Fock states of the many mode field by $|\{n_j\}\rangle
\equiv |n_1,n_2,...\rangle$. Because the present discussion concerns the
external system interacting the considered system , the conceptions
presented in last section must be alternated. Actually, in stead of the
effective ``two-time wave function'', we use the effective two-time state
vector
\begin{equation}
|\psi _B(t,t^{\prime })\rangle =\hat{B}(t^{\prime })\hat{B}(t)|\psi
(0)\rangle .
\end{equation}
Then we can re-write the second order correlation function as
\begin{equation}
G[t,t^{\prime },\hat{\rho}(0)]=\langle \psi _B(t,t^{\prime })|\psi
_B(t,t^{\prime })\rangle
\end{equation}

It is interesting that the effective state vector can be evaluated as the
superposition
\begin{widetext}
\begin{equation}
|\psi _B(t,t^{\prime })\rangle =\frac 12e^{i\hat{V}(0,0)t^{\prime
}}[\exp (-i\omega _Vt^{\prime })e^{-i\hat{V}(1,0)t^{\prime
}}e^{i\hat{V}(1,0)t}+ \exp (-i\omega _Vt)e^{i\hat{V}(0,0)t^{\prime
}}e^{-i\hat{V}(0,1)t^{\prime
}}e^{i\hat{V}(0,1)t}]e^{-i\hat{V}(1,1)t}|\{0_j\}\rangle \otimes
|0_H,0_V\rangle
\end{equation}
\end{widetext}
of two components for the two paths from the initial two particle state $%
|1_H,1_V\rangle $ to the two particle vacuum $|0_H,0_V\rangle $ . It should
be noticed that the effective actions of the reservoir
\begin{widetext}
\begin{equation}
\hat{V}(m,n)\equiv \sum_j\hat{V}_j(m,n)=\sum_j\omega _j\hat{a}_j^{\dagger }%
\hat{a}_j+\sum_j(d_V(\omega _j)m+d_H(\omega _j)n)(\hat{a}_j^{\dagger }+\hat{a%
}_j)
\end{equation}
\end{widetext}
can label the different paths and record the path information in the
reservior. Thus , this generalized ''which-path measurement '' leads to the
second order quantum decoherence.

The above result clearly demonstrates that, with the presence of the
reservoir, the different probability amplitudes ($\sim $ $\exp (-i\omega
_Vt^{\prime })$ and $\exp (-i\omega _Vt)$) from $|1_H,1_V\rangle $ to $%
|0_H,0_V\rangle $ entangle with the different states
\[
(\frac 12e^{i\hat{V}(0,0)t^{\prime }}e^{-i\hat{V}(1,0)t^{\prime }}e^{i\hat{V}%
(1,0)t}e^{-i\hat{V}(1,1)t}|\{0_j\}\rangle
\]
and
\[
\frac 12e^{i\hat{V}(0,0)t^{\prime }}e^{-i\hat{V}(0,1)t^{\prime }}e^{i\hat{V}%
(0,1)t}e^{-i\hat{V}(1,1)t}|\{0_j\}\rangle)
\]
of the reservoir. This is just the physical cause of the second
order quantum decoherence. In the following section an explicit
calculation of the second order correlation function will be given
to illustrate this crucial observation.

\section{Dynamic Decoherence in Higher Order Case}

After a straightforward calculation the second order correlation function
can be expressed in a factorization form [9]:
\begin{equation}
G[t,t^{\prime },\hat{\rho}(0)]=\frac 12[1+{{\mathbf Im}(e^{i\omega
_V(t-t^{\prime })}\prod_jF_j)]}  \label{eqG}
\end{equation}
where each factor
\begin{widetext}
\begin{equation}
F_j =\langle 0_j|e^{i\hat{V}_j(1,1)t}e^{-i\hat{V}_j(0,1)t}e^{i\hat{V}%
_j(0,1)t^{\prime }}e^{-i\hat{V}_j(1,0)t^{\prime }}e^{i\hat{V}_j(1,0)t}e^{-i%
\hat{V}_j(1,1)t}|0_j\rangle  \equiv \langle
0_j|\hat{u}_j^6(t_6)|0_j\rangle
\end{equation}
\end{widetext}
is a two-time transition amplitude of the $j^{\prime }th$ mode of the
reservoir. Obviously, the term $\prod_jF_j$ measures the extent of coherence
and decoherence in the second order case. It plays the same role as the
decoherence factor of the first order decoherence \cite{Sun-qm} . So it is
also called the decoherence factor.

In the following, to give the factor $F_j$ explicitly, we adopt
the Wei-Norman method \cite{Wei,Sun-xiao} to calculate the
effective time evolution defined by $\hat{u}_j^6(t_6)$. It can be
imagined as an evolution
governed by a discrete time-dependent Hamiltonian $H(t)$ dominated by $\hat{V%
}_j(1,1),-\hat{V}_j(1,0),\hat{V}_j(1,0),-\hat{V}_j(0,1),\hat{V}_j(0,1)$ and $%
-\hat{V}_j(1,1)$ in six time-intervals $%
[t_0=0,t_1=t],[t_1,t_2=2t],[t_2,t_3=2t+t^{\prime }],[t_3,t_4=2t+2t^{\prime
}],[t_4,t_5=3t+2t^{\prime }],$ $[t_5,t_6=4t+2t^{\prime }]$ respectively. In
the $k$-th step of calculation, we take the final state of ($k-1)$-th step
as its initial state. Therefore, we obtain $\hat{u}_j^6(t_6)$ as the sixth
step evolution

\begin{equation}
\hat{u}_j^6(t_6)=e^{{g_{1j}^6(t_6)\hat{a}_j}^{\dagger }}e^{{g_{2j}^6(t_6)%
\hat{a}_j}^{\dagger }\hat{a}_j}e^{{g_{3j}^6(t_6)}\hat{a}_j}e^{{g_{4j}^6(t_6)}%
}
\end{equation}
Here, ${g_{kj}^6(t_6)(k=1,2,3,4)}$ are the coefficients that can be
explicitly obtained. But for the calculation of the $j$-th component
\begin{equation}
F_j=\exp [{g_{4j}^6}(t_6)]  \label{eqF}
\end{equation}
of the decoherence factor$,$ we only need to know ${g_{4j}^6.}$ The detailed
discussion in the appendix gives
\begin{widetext}
\begin{eqnarray}
{g_{4j}^6}(t_6) &=&-\frac 2{\omega _j^2}[d_H(j)-d_V(j)]^2\sin ^2[\frac
12\omega _j(t^{\prime }-t)]+\frac i{\omega _j^2}[d_V^2(j)-d_H^2(j)]
\nonumber \\
&\mbox{}&[\omega _j(t^{\prime }-t)+2(1-\cos (\omega _j[t^{\prime }-t])\sin
\omega _jt]+(1-2\cos \omega _jt)\sin (\omega _j[t^{\prime }-t])].  \nonumber
\\
&=&-R_j(t-t^{\prime })+i\Omega _j(t,t^{\prime })  \label{equ}
\end{eqnarray}
\end{widetext}
It is noticed that the real part

\[
-R_j(t-t^{\prime })=-\frac 2{\omega _j^2}[d_H(j)-d_V(j)]^2\sin ^2[\frac
12\omega _j(t^{\prime }-t)]
\]
only depends on the time interval $t^{\prime }-t,$ but the imaginary part $%
\Omega _j(t,t^{\prime })$ depends on both $t^{\prime }$ and $t^{\prime }$ as
a two time function.

Because $-R_j$ can not exceed zero, the norm $|F_j|=e^{-R_j(t)}$of the
factor $F_j$ can not exceed one . Then from the arguments about the first
order decoherence in our previous works on quantum measurement theory \cite
{Sun-qm}, it is concluded qualitatively that the factorization structure of
the decoherence factor
\begin{eqnarray*}
F &=&\prod_{j=1}^NF_j=\prod_{j=1}^Ne^{i\Omega _j(t,t^{\prime })}\cdot
\prod_{j=1}^Ne^{-R_j(t-t^{\prime })} \\
&=&|F|\exp [i\Omega (t,t^{\prime })]
\end{eqnarray*}
implies the vanishment of the second order correlation in the macroscopic
limit that the number $N$ of particles making up the reservoir approaches
the infinity. This is because
\[
|F|=\exp [-\sum_{j=1}^NR_j(t,t^{\prime })]\rightarrow 0
\]
as $N\rightarrow \infty $ since $\sum_{j=1}^NR_j(t,t^{\prime })$ is a
diverging series or a monotonously-increasing function of $t-t^{\prime }$
under some reasonable conditions.

In order to demonstrate the above conclusion quantitatively, we give the
numerical results for the second order decoherence for different numbers N
of the quantum oscillators. As $N$ increase, these results are illustrated
in Fig.1. In the numerical calculation, the coupling constants $\{d_V(j)\}$
take random values in the domain $[0.8,1.0]$, the coupling constants $%
\{d_H(j)\}$ in $[0.2,0.4]$, and the frequencies $\{\omega _j\}$ in $%
[0.5,1.5] $. The other parameters are given in the caption of the figure. %
\vskip 0.2cm
\begin{figure}[tbp]
\begin{tabbing}
\= \includegraphics[width=8cm]{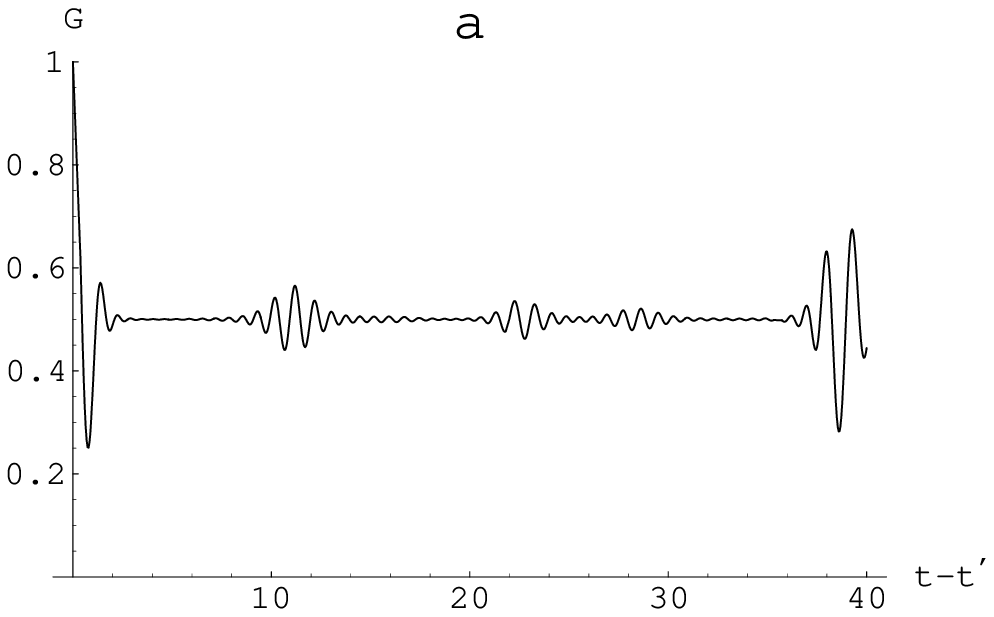}\\  \>
\includegraphics[width=8cm]{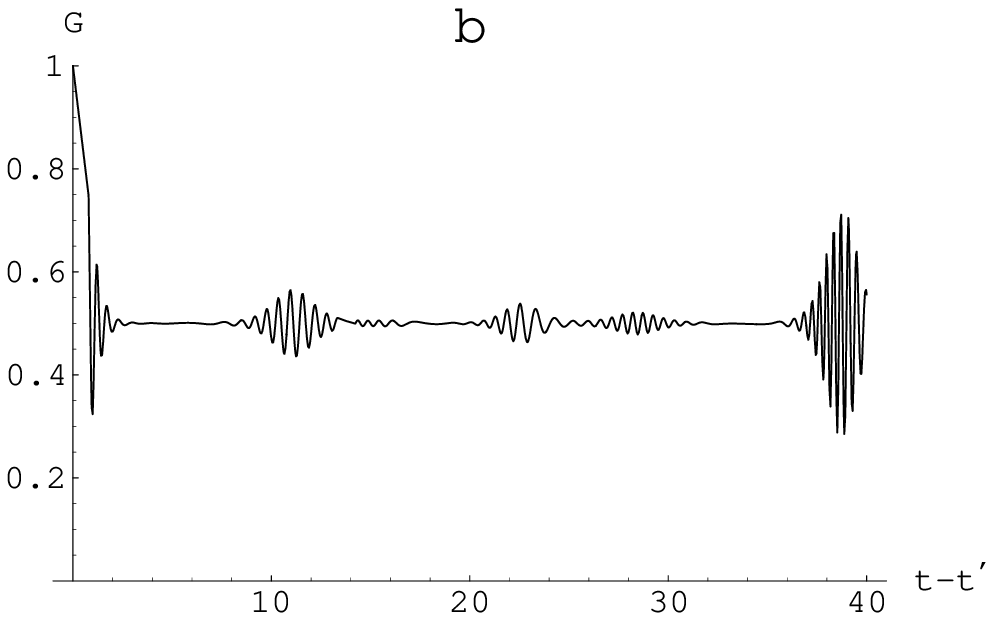}\\  \>
\includegraphics[width=8cm]{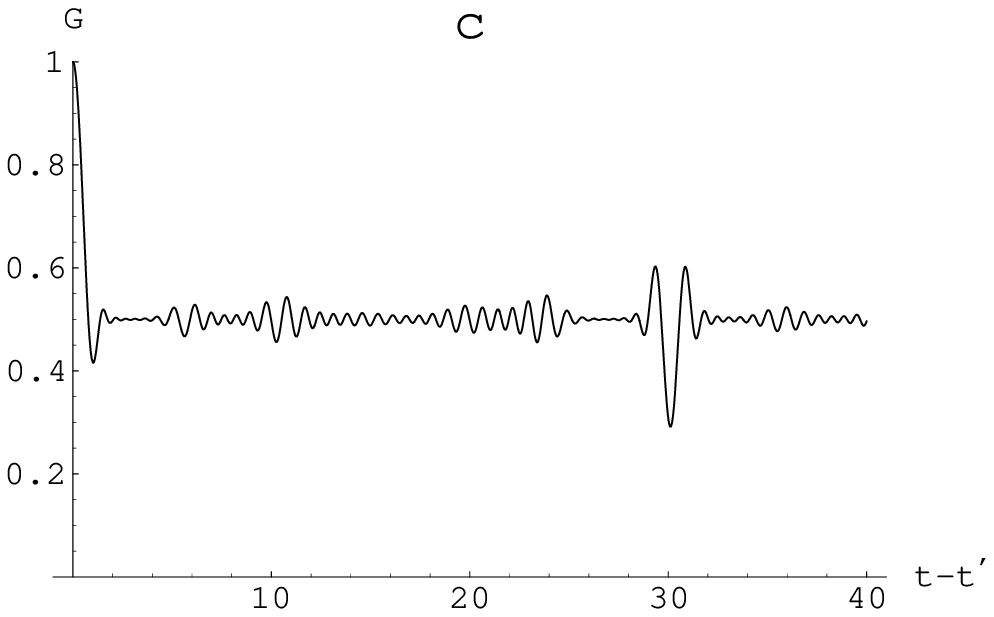} \\ \>
\includegraphics[width=8cm]{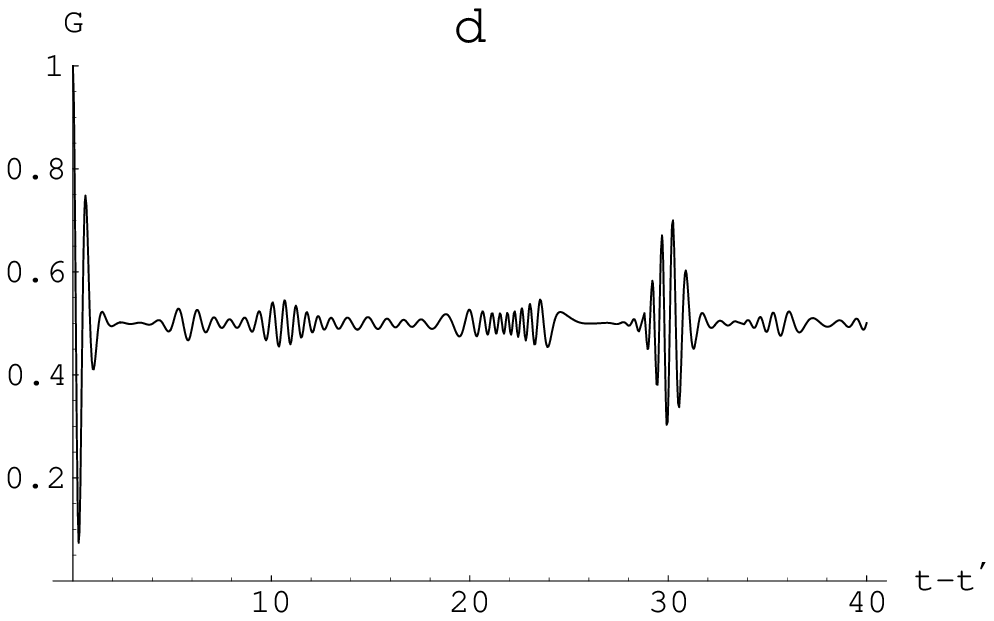}
\end{tabbing}
\vskip 0.2cm
\caption{The horizontal axe denotes time period $t^{\prime }-t$, the
vertical axe denotes the second order correlation function $G\equiv
G[t,t^{\prime },\hat{\rho}(0)]$, parameters ${\omega _V=1.0}$, (a)${N=5,t=0}$%
, (b)${N=5,t=5}$, (c)${N=10,t=0}$, (d)${N=10,t=5}$.}
\end{figure}

From both the above graphic illustration given by the numerical
calculation and the analytic result given by
Eqs.(\ref{eqG},\ref{eqF},\ref{equ}) , we see clearly that the
second order correlation function depends not only on the time
interval $t^{\prime }-t$, but also on $t$ or $t^{\prime }$. In
fact it follows from the analytic result
\begin{equation}
{{\mathbf Im}(e^{i\omega _V(t-t^{\prime })}\prod_jF_j)=\frac 12|F|\cos [\omega
_V(t-t^{\prime })+\Omega (t,t^{\prime })]}
\end{equation}
that the amplitude of the second order correlation function is mainly
determined by the time interval $t^{\prime }-t$, but the phase is determined
by the two time parameters . What is more important is that as the number of
the quantum oscillators increases, the second order coherence vanishes
faster and faster, and the amplitude for quantum revival becomes smaller and
smaller. With a reasonable extrapolation, it can be predicted that, when the
number of the quantum oscillators approaches the infinity in the macroscopic
limit, the second order coherence will vanish in a very short time and thus
no quantum revival phenomenon can be observed.

\section{Concluding Remarks}

In this paper we first depict the high order quantum coherence of a boson
system by introducing the concept of multi-particle wave amplitude. For some
cases with the specifically- given initial state , we show that the norm
square of multi-particle wave amplitude (or a sum of the norm squares for an
open system) gives the high order correlation function. As an effective
multi-time wave function, this amplitude can be shown to be a superposition
of several ``multi-particle paths''. When the environment or an apparatus
entangles with them to form a generalized ``which-path(way)'' measurement,
the high order quantum decoherence happens dynamically. Some explicit and
general illustrations are presented in this paper to construe our
observation. But to prove our conjecture that any high order correlation
function indicating the existence of high order quantum coherence can be
expressed as the norm square of a properly-defined effective wave amplitude
for many particle system, there is still a long way to go. If this
conjecture is true under certain general conditions, then there still exist
the problem of clarifying these conditions. Moreover, experimental proposals
that can be implemented at least in principle are still unavailable.

Our present investigations shed a new light on the understanding of quantum
coherence . According to arguments in this paper the high order quantum
coherence can also be sculptured as the generalized interference phenomenon
by two ``multi-particle paths '', and the intrinsically quantum features
(higher order ones) of coherence beyond the classical analogue reflected by
the spatial interference of two paths in classical electromagnetic field can
be theoretically unified in a framework to embody the wave - particle
duality in the quantum world. Therefore, as the essential element, the
''which-path(way)'' detection in both the original and the extended
versions, naturally provides a complete decoherence mechanism in
understanding quantum measurement and the transition from quantum to
classical mechanics.

\begin{acknowledgements}
This work is supported by the
NSF of China and the knowledged Innovation Programme(KIP) of the
Chinese Academy of Science. X.F.Liu is  acknowledged for his
kindly patients in reading  and writing this manuscript. The
authors also thanks the referee for his  some useful suggestions
to improve the presentation  of this paper.
\end{acknowledgements}

\appendix

\section*{Appendixes}

In this appendix, the Wei-Norman method \cite{Wei,Sun-xiao} is adopted to
calculate the second order decoherence factor $F_j$. The calculation is
completed in six steps.

During the time period $[t_{k-1},t_k](k=1,2,\cdots ,6)$, Let $%
W_j^k(t,t^{\prime })$ be an time evolution dominated by the single particle
Hamiltonian
\begin{equation}
\hat{h}_j^k={\alpha _j^k}\hat{a}_j^{\dagger }\hat{a}_j+{\beta _j^k}\hat{a}%
_j^{\dagger }+{\gamma _j^k}\hat{a}_j,\hspace{0.5cm}\{k=1,2,\cdots ,6\}.
\end{equation}
The coefficients $\alpha _j^k,\beta _j^k,\gamma _j^k$ and the time intervals
$T_k=t_k-t_{k-1}$ take different values in the six different steps:
\begin{eqnarray}
&&\alpha _j^1=\omega _j,\beta _j^1=\gamma _j^1=d_V(\omega _j)+d_H(\omega
_j),T_1=t,  \nonumber \\
&&\alpha _j^2=-\omega _j,\beta _j^2=\gamma _j^2=d_V(\omega _j),T_2=t,
\nonumber \\
&&\alpha _j^3=\omega _j,\beta _j^3=\gamma _j^3=d_V(\omega _j),T_3=t^{\prime
},  \nonumber \\
&&\alpha _j^4=-\omega _j,\beta _j^4=\gamma _j^4=-d_H(\omega
_j),T_4=t^{\prime },  \nonumber \\
&&\alpha _j^5=\omega _j,\beta _j^5=\gamma _j^5=d_H(\omega _j),T_5=t,
\nonumber \\
&&\alpha _j^6=\omega _j,\beta _j^6=\gamma _j^6=-d_V(\omega _j)-d_H(\omega
_j),T_6=t.
\end{eqnarray}
Because of the fact that the four operators ${n}_j=\hat{a}_j^{\dagger }%
\hat{a}_j$, $\hat{a}_j^{\dagger }$, $\hat{a}_j$ and $1$ form a closed
algebra - the Heisenberg-Wely algebra, the unitary time evolution operator
at each step takes the following form (Wei-Norman theorem)
\begin{equation}
\hat{u}_j^k(T)=e^{{g_{1j}^k}(T){\hat{a}_j}^{\dagger }}e^{{g_{2j}^k}(T){%
\hat{a}_j}^{\dagger }\hat{a}_j}e^{{g_{3j}^k}(T)\hat{a}_j}e^{{g_{4j}^k}(T)}.
\end{equation}
for $T\in [t_{k-1},t_k]$ in a special sequence. Here coefficients ${g_{sj}^k}%
(T)(s=1,2,3,4)$ are  functions of $T$ to be determined. The benefit of the
above form is that only the coefficient ${g_{4j}^k}(T)$ is needed  in the
calculation of the average value at the vacuum state. So we can largely
reduce the complexity of our calculation as we need to  pay  attention only
to things concerning ${g_{4j}^k}(T)$ .

Substituting $\hat{u}_j^k(T)$ into the Schr$\ddot{o}$dinger equation
\[
i\frac d{dT}\hat{u}_j^k=\hat{h}_j^k\hat{u}_j^k,
\]
we find the coefficients ${g_{sj}^k}(T)(s=1,2,3,4)$ satisfy the following
system of equations:
\begin{eqnarray}
\frac d{dT}{g_{2j}^k} &=&-i{\alpha _j^k},  \nonumber \\
\frac d{dT}{g_{1j}^k}-{g_{1j}^k}\frac d{dT}{g_{2j}^k} &=&-i{\beta _j^k},
\nonumber \\
e^{-{g_{2j}^k}}\frac d{dT}{g_{3j}^k} &=&-i{\gamma _j^k} \\
\frac d{dT}{g_{4j}^k}-{g_{1j}^k}e^{-{g_{2j}^k}}\frac d{dT}{g_{3j}^k} &=&0
\nonumber
\end{eqnarray}
Using the results
\begin{equation}
\frac d{dT}{g_{1j}^k}=-i{\alpha _j^k}{g_{1j}^k}-i{\beta _j^k},\frac d{dT}{%
g_{4j}^k}=-i{\gamma _j^k}{g_{1j}^k}
\end{equation}
obtained by simplifying the above system of equations , we get the solution
\begin{eqnarray}
{g_{1j}^k}(T) &=&({g_{1j}^k}(t_{k-1})+\frac{\beta _j^k}{\alpha ^k})e^{-i{%
\alpha _j^k}(T-t_{k-1})}-\frac{\beta _j^k}{\alpha _j^k},  \nonumber \\
{g_{4j}^k}(T) &=&{g_{4j}^k}(t_{k-1})+\frac{\gamma _j^k}{\alpha _j^k}({%
g_{1j}^k}(t_{k-1})+\frac{\beta _j^k}{\alpha _j^k})\nonumber\\
&&(e^{-i{\alpha
_j^k}(T-t_{k-1})}-1)+i\frac{{\beta _j^k}{\gamma _j^k}(T-t_{k-1})}{\alpha _j^k}
\end{eqnarray}
Notice that, to obtain the above result we have used the initial conditions
\begin{eqnarray}
g_{1j}^k(t_{k-1}) &=&g_{1j}^{k-1}(t_{k-1}), \\
g_{4j}^k(t_{k-1}) &=&g_{4j}^{k-1}(t_{k-1})
\end{eqnarray}
for each step and the initial conditions $g_{1j}^0(t_0)=g_{4j}^0(t_0)=0$ for
the first step. Then, we obtain a set of iteration equations
\begin{eqnarray}
g_{1j}^k(t_k) &=&({g_{1j}^{k-1}}(t_{k-1})+\frac{\beta _j^k}{\alpha _j^k}%
)e^{-i{\alpha ^k}T_k}-\frac{\beta _j^k}{\alpha _j^k},  \nonumber \\
g_{4j}^k(t_k) &=&{g_{4j}^{k-1}(t_k-1)}+\frac{\gamma _j^k}{\alpha _j^k}({%
g_{1j}^{k-1}(t_{k-1}}+\frac{\beta _j^k}{\alpha _j^k})\nonumber\\
&&(e^{-i{\alpha _j^k}%
T_k}-1)+i\frac{{\beta _j^k}{\gamma _j^k}T_k}{\alpha ^k}. \label{eq2}
\end{eqnarray}

Iterating six times with different initial conditions and coefficients, the
final result of ${g_4^6(t_6)}$ is obtained as Eq.(\ref{eq2}).



\begin{thebibliography}{}
\bibitem{Glauber}  R.J. Glauber, Phys.Rev.{\textbf 130}, 2529(1963); {\textbf 131}, 2766(1963).
\bibitem{Twiss}  H. Hanburg-Brow and R.Q. Twiss, Phil.Mag.{\textbf
45},663(1954); Nature{\textbf 178}, 1046(1956); Proc.Roy.Soc.A{\textbf
242},300(1957).
\bibitem{Rampe}  S. Durt, T. Nonn, and G. Rampe, Nature {\textbf 395}, 33(1998).
\bibitem{Umansky}  E. Buks, R. Schuster, M. Heiblum, D. Mahalu, and V.
Umansky, Nature {\textbf 391}, 871(1998).
\bibitem{Zurek}  W.H. Zurek,
Phys.Today, {\textbf 44(10)}, 36 (1991).
\bibitem{Haroche}  M. Brune, E. Hagley, J. Dreyer, X. Maitre, A. Maali, C.
Wunderlich, J. M. Raimond, and S. Haroche. Phys. Rev. Lett. {\textbf
77}, 4887(1996).
\bibitem{Sun-qm} C. P. Sun, Phys. Rev. A 48, 878(1993). C. P. Sun,
Chin. J. Phys. {\textbf 32}, 7(1994). C. P. Sun, X. X. Yi, and X. J.
Liu, Fortschr. Phys. {\textbf 43}, 585.
\bibitem{Knight} S. Bose, K. Jacobs, P. Knight, Phys.Rev. A {\bf
82}, 3204(1999).
\bibitem{Unruh} W.Unruh,Phys.Rev. A {\textbf51}, 992(1995).
\bibitem{Palma} G. Palma, K. Suominen, A. Ekert, {\textsl Proc. Roy. Soc.
London} Ser. A {\textbf  452}, 567 (1996)
\bibitem{Sun-qc}C. P. Sun, H. Zhan, and X. F.
Liu, Phys. Rev. A {\textbf 58}, 1810(1998).
\bibitem{Sun-adia} C. P. Sun,X. F. Liu, D. L. Zhou, and S. X.
Yu, Phys.Rev. A {\textbf63}, 2111(2001), C. P. Sun, D. L. Zhou, S. X.
Yu, and X. F. Liu, Eur. J. Phys. D {\textbf 13}, 145(2001);C. P.
Sun,X. F. Liu, D. L. Zhou, and S. X. Yu, Eur. J. Phys. D {\textbf 17},
89(2001)
\bibitem{Zhou-dl} D.L. Zhou, and C.P. Sun, quant-ph/0104038, LNAL
preprint(2001).
\bibitem{Scully} M.O. Scully, and M.S. Zubairy, {\textit Quantum
Optics}, Cambridge University press, 1997, pp 97-129.
\bibitem{Shih} Y. Shih, Adv.At.Mol.Opt.Phys. Vol 41(1999),
pp 1-42.
\bibitem{footnote} This definition, which is defined by Klauder and Sudarshan in
{\it Fundamentals of Quantum Optics}, W.A. Benjamin, NewYork,
1968, more convenient to describe coherence as interference, has a
little difference in the denominator with Glauber's one. Glauber's
original definition is
\begin{widetext}
\[
g[\alpha _1,\alpha _2,\cdots ,\alpha _n;\beta _1,\beta _2,\cdots ,\beta _n]
=
\frac{Tr[\hat{\rho}B_{\beta _1}^{\dagger }(t_1)B_{\beta
_2}^{\dagger }(t_2)\cdots B_{\beta _n}^{\dagger }(t_n)B_{\alpha
_n}(t_n)\cdots B_{\alpha _2}(t_2)B_{\alpha
_1}(t_1)]}{\sqrt{G[\alpha _1,t_1]G[\alpha _2,t_2]\cdots G[\alpha
_n,t_n]}\sqrt{G[\beta _1,t_1]G[\beta _2,t_2]\cdots G[\beta
_n,t_n]}}
\]
\end{widetext}
\bibitem{Wei} J. Wei and E. Norman, J. Math.Phys. {\textbf 4A},
575(1963).
\bibitem{Sun-xiao} C.P. Sun and Q. Xiao, Commun.Theor.Phys. {\textbf 16},
359(1990).
\end{thebibliography}
\end{document}